\documentclass[twocolumn,showpacs,showkeys,preprintnumbers,superscriptaddress]{revtex4-1}
\usepackage{amssymb}

\usepackage{amsfonts}

\usepackage{latexsym}

\usepackage{dcolumn}

\usepackage{amsmath}

\usepackage{graphicx}
\usepackage{psfrag,pstricks}

\newcommand{\rmi}{\mathrm{i}}

\newcommand{\rmm}{\mathrm{m}}

\newcommand{\rmin}{\mathrm{in}}

\begin{document}

\title{Steady-state entanglement and normal-mode splitting in an atom-assisted optomechanical system with intensity-dependent coupling }
\author{Sh. Barzanjeh}
\email{shabirbarzanjeh@yahoo.com}
\affiliation{Department of Physics, Faculty of Science,
University of Isfahan, Hezar Jerib, 81746-73441, Isfahan, Iran}
\affiliation{School of Science and Technology, Physics Division, Universit\`{a} di Camerino,
I-62032 Camerino (MC), Italy}
\author{M. H. Naderi}
\email{mhnaderi@phys.ui.ac.ir}
\author{ M. Soltanolkotabi}
\email{soltan@sci.ui.ac.ir}
\affiliation{Quantum Optics Group, Department of Physics, Faculty of Science, University of Isfahan, Hezar Jerib, 81746-73441, Isfahan, Iran}
\date{\today}

\begin{abstract}

In this paper, we study theoretically the bipartite and tripartite continuous variable entanglement as well as the normal-mode splitting in a single-atom cavity optomechanical system with intensity-dependent coupling. The system under consideration is formed by a Fabry-Perot
cavity with a thin vibrating end mirror and a two-level atom in the Gaussian standing-wave of the cavity mode. We first derive the general form of Hamiltonian describing the tripartite intensity-dependent atom-field-mirror coupling due to the presence of cavity mode structure. We then restrict our treatment to the first vibrational sideband of the mechanical resonator and derive a novel form of tripartite atom-field-mirror Hamiltonian. We show that when the optical cavity is intensely driven one can generate bipartite entanglement between any pair of the tripartite system, and that, due to entanglement sharing, the atom-mirror entanglement is efficiently generated at the expense of optical-mechanical and optical-atom entanglement. We also find that in such a system, when the Lamb-Dicke parameter is large enough one can simultaneously observe the normal mode splitting into three modes.
\end{abstract}


\maketitle

\section{Introduction}
%
%
Cavity optomechanics is a rapidly growing field of research that is concerned with the interaction between a mechanical resonator (MR) and the radiation pressure of an optical cavity field\cite{cave,corbitt1,vitali1,vit,kipp,fabre}. The optomechanical coupling widely employed
for a large variety of applications\cite{cohadon}, more commonly as a sensor for the detection of weak forces\cite{brad} and small displacements\cite{laha} or an actuator in
integrated electrical, optical, and opto-electronical systems\cite{tang,brown}. However, the most experimental and theoretical efforts are devoted to cooling and trapping such mechanical resonators to their quantum ground state, which more recently have been done successfully\cite{teu}. Furthermore, in Ref.\cite{nori11} the authors have proposed a different scheme to enhance the cooling process by using the photothermal (bolometric) force\cite{nori22}. They have taken into account the noise effects due to the granular nature of photon absorption and finally have shown that the mechanical resonator can achieve the lowest phonon occupation number by means of this procedure. Moreover, it seems promising for the realization of long-range interaction between qubits in future quantum information hardwares \cite{rabl}, and for probing quantum mechanics at increasingly large mass and length scales\cite{Marshall}. The coupling of a MR via radiation pressure to a cavity field shows interesting similarities to an intracavity nonlinear Kerr-like interaction\cite{fabre} or even a more complicated form of nonlinearity\cite{barzanjeh}.

To observe and control quantum behavior in an optomechanical system, it is essential to increase the strength of the coupling between the mechanical and optical degree of freedom. However, the form of this coupling(e.g., linear or nonlinear) is crucial in determining which phenomena can be observed in such a system. Thanks to rapid progress of nano-technology, it has been possible to manipulate the optomechanical coupling in quantum optomechanical hybrid systems. In this direction most
experimental and theoretical efforts are devoted to entangling a MR either with a single atom \cite{atom-membrane,Yue,Hammerer2009,wang,camer} or with atomic ensembles \cite{Reichel07,genes,ian-ham,ham,treutlein,kanamoto}, entangling a nanomechanical oscillator with a Cooper-pair box \cite{Armour03}, and entangling two charge qubits \cite{zou1} or two Josephson junctions \cite{cleland1} via nanomechanical resonators. Alternatively, schemes for entangling a superconducting coplanar waveguide field with a nanomechanical resonator, either via a Cooper-pair box within the waveguide \cite{ringsmuth}, or via direct capacitive coupling \cite{Vitali07}, have been proposed.

In Ref.\cite{genes} the authors have proposed a scheme for the realization of a hybrid, strongly quantum-correlated system consisting of an atomic ensemble surrounded by a high-finesse optical cavity with a vibrating mirror. They have shown that, in an experimentally accessible parameter regime, the steady state of the system shows both tripartite and bipartite continuous variable(CV) entanglement.
More recently, the dynamics of a movable mirror of a cavity coupled through radiation pressure to
the light scattered from ultracold atoms in an optical lattice has been investigated Ref.\cite{bhata}. The author has shown that in the presence of atom-atom interaction as a source of nonlinearity\cite{haghshenas}, the coupling of the mechanical oscillator, the cavity field fluctuations and the condensate fluctuations (Bogoliubov mode) leads to the splitting of the normal mode into three modes (normal-mode splitting(NMS)\cite{sanch,agarwal1984,raizen,boca,maunz}). The system described there shows a complex interplay between distinctly three systems namely, the nanomechanical cantilever, optical microcavity and the gas of ultracold atoms.

The optomechanical NMS is one of the fascinating phenomena arising from the
strong coupling between the cavity and the mechanical mirror\citep{sumei,dob,tar}. In Ref.\cite{dob} it has been shown that the cooling of mechanical oscillators in the resolved sideband regime at high driving power laser can entail the appearance of NMS. Moreover,  the dynamics of a movable mirror of a nonlinear optical cavity is considered in Ref.\cite{tar}. It has been shown that a $\chi^{(3)}$
medium with a strong Kerr nonlinearity placed inside the cavity inhibits the NMS due to the photon blockade mechanism(this just happens only if the Kerr nonlinearity is much greater than the cavity decay rate).  As the authors have shown in Refs.\cite{bhata} and \cite{tar} the nonlinearity plays a crucial role in the appearance of NMS in the optomechnical systems. 

The main purpose of the present paper is to study the quantum behavior of an atom-assisted cavity optomechanical system in which a single two-level atom is trapped in the standing-wave light field of a single-port Fabry-Perot cavity. The infinite set of optical modes of the cavity can be described by Hermit-Gauss modes. As we will see, the intracavity mode structure can be employed to realize a type of intensity-dependent coupling of the single atom to the vibrational mode of MR.
The presence of such intensity-dependent interaction modifies the dynamics of the system, the entanglement properties and the displacement spectrum of MR. We show that in the first vibrational sideband of MR,  the stationary, i.e., long-lived, atom-mirror entanglement can be generated by proper matching the Lamb-Dicke parameter(LDP). This parameter plays an important role in our investigation in the sense that it determines the strength of the nonlinearity in the system. We show that the bipartite entanglement between the subsystems extremely depends on the LDP. It is also remarkable that, in the
steady-state condition, the high resolution of NMS in the form of three-mode splitting is approached. In particular, the appearance of intensity-dependent coupling leads to a progressive increasing of NMS due to the strong nonlinear atom-field-mirror interaction.

The paper is organized as follows. In Sec. II we derive an intensity-dependent
Hamiltonian describing the triple coupling of atom-field-mirror through  \textit{j}-phonon excitations of the vibrational sideband. In Sec. III, we derive the quantum
Langevin equations (QLEs) and linearize them around the semiclassical steady
state. In Sec. IV we study the steady state of the system and quantify the entanglement properties of the system by using the logarithmic negativity. In Sec. V we investigate the appearance of NMS in the displacement spectrum of the mirror. Our conclusions are summarised in Sec. VI.
\section{Model}
The system studied in this paper is sketched in Fig.~1. It consists of a hybrid system formed by a single two-level atom with transition frequency $\omega_{\mathrm{e}}$ which
trapped in the standing-wave light field of a single-port Fabry-Perot cavity with a movable
mirror coated on the plane side of a mechanical resonator. The geometry of the resonator determines the spatial
structure of the acoustic modes.
The movable mirror is treated as a quantum mechanical harmonic oscillator with effective mass $m$, frequency $\omega_{\mathrm{m}}$,
and energy decay rate $\gamma_{\mathrm{m}}$. The system is also coherently driven by a laser field with frequency $\omega_{l}$ through the
cavity mirror with amplitude $ \mathcal{E}$. We assume that the single atom is indirectly coupled to the mechanical oscillator via the
common interaction with the intracavity field with frequency $\omega_{\mathrm{c}}$.
\begin{figure}[ht]
\centering
\includegraphics[width=3.8in]{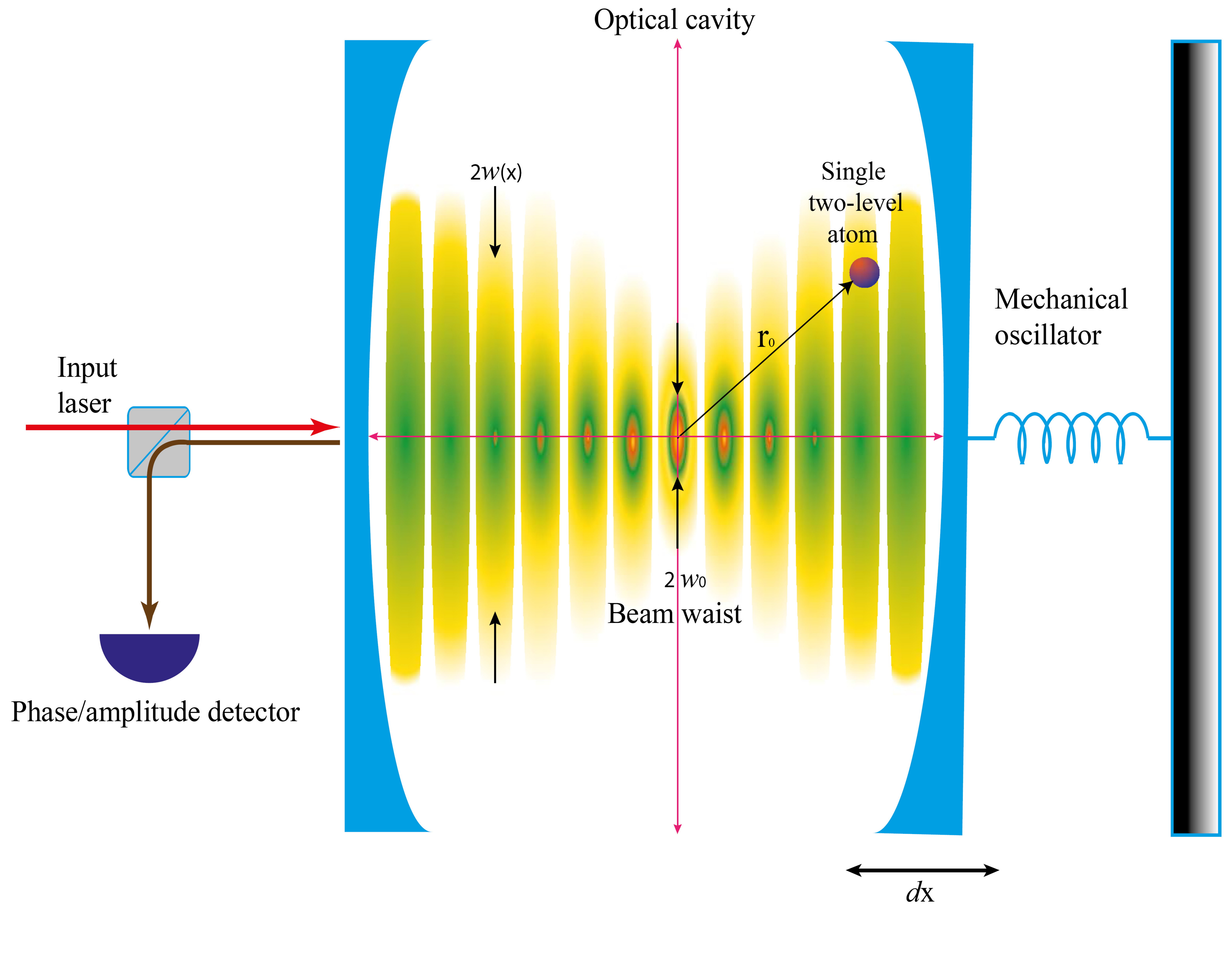}
\caption{The schematic of the atom-assisted optomechanical system. It contains an optical cavity ended with a fixed mirror and a slightly moving mirror which is attached to a spring. Inside the cavity there is a two-level atom. The system is coherently driven by a laser field}
\end{figure}

In our investigation we can restrict the model to the case of single-cavity and mechanical modes. This
is justified when the cavity free spectral range is much larger
than the mechanical frequency $\omega_{\mathrm{m}}$(i.e., not too large cavities).
In this case, scattering of photons from the driven mode into other cavity modes is negligible \cite{law1995} and the input
laser successfully drives only one cavity mode.
This guarantees the fact that only one cavity mode participates in the optomechanical interaction and the
neighbouring modes are not excited by a single central frequency input laser.
In addition, one can restrict to a single
mechanical mode when the detection bandwidth is chosen such
that it includes only a single, isolated, mechanical resonance
and the mode-mode coupling is negligible \cite{Genes2008}.
\subsection{Hamiltonian of the system}
In the absence of dissipation and fluctuations,
the total Hamiltonian of the system is given by the
sum of three term; the free evolution term\cite{Yue,genes}
\begin{eqnarray}\label{free}
H_0&=&\hbar\omega_{\mathrm{c}}a^{\dagger}a+\hbar\omega_{\mathrm{m}}b^{\dagger}b+\frac{\hbar\omega_{\mathrm{e}}}{2}\sigma^{\mathrm{z}},
\end{eqnarray}
the interaction term
\begin{equation}
H_{\mathrm{int}}=-\hbar \xi_0(b+b^\dagger)a^{\dagger}a+\hbar\chi_{mnl}(\vec r_0,x)[a\sigma^{+}+h.c\big],
\label{hamiltonian1}
\end{equation}
and the laser driven term
\begin{eqnarray}
H_{\mathrm{dri}}=\mathrm{i}\hbar \mathcal{E}(a^{\dagger}\mathrm{e}^{-\mathrm{i}\omega_{l}t}-ae^{\mathrm{i}\omega_{l}t}),
\label{hamiltonian2}
\end{eqnarray}
where $a$($[a,a^{\dag}]=1$) is the annihilation operator of the cavity field with the decay
rate $\kappa$, $b$($[b,b^{\dag}]=1$) is the motional annihilation operator of the MR, and the single two-level
atom is described by the
spin-$1/2$ algebra of the Pauli matrices $\sigma^{-}$, $\sigma^{+}$ and $\sigma^{\mathrm{z}}$ which satisfy
the commutation relations$[\sigma^{+},\sigma^{-}]=\sigma^{\mathrm{z}}$ and $[\sigma^{\mathrm{z}},\sigma^{\pm}]=\pm2\sigma^{\pm}$. It should be noted that the free Hamiltonian(\ref{free}) has been written within the \textit{Raman-Nath} approximation\cite{Schleich}, i.e., in the limit when the atom is allowed only to move over a distance which is much less than wavelength of the light. Therefore, in this approximation, one can neglect the kinetic energy of the atom. The first term of $H_{\mathrm{int}}$
is the optomechanical coupling with the radiation-pressure coupling constant $\xi_0=(\omega_{\mathrm{c}}/L)x_{\mathrm{ZPF}}$, in which $x_{\mathrm{ZPF}}=\sqrt{\hbar/m \omega_{\mathrm{m}}}$
is the zero point fluctuations of mechanical oscillator. The second term of $H_{\mathrm{int}}$ denotes the "three-body" interactions
among the atom, the cavity field and the vibration of the mirror. The field-atom coupling rate in terms of an infinite set of optical modes is well described by the Hermite-Gauss modes\cite{Siegman,vitali2011}
\begin{equation}\label{coupling1}
\chi_{mnl}(\vec r)=g_0K_{mnl}(x,y,z){\mathrm{sin}}\Big[\psi_{mnl}(x,y,z)-\frac{l\pi}{2}\Big],
\end{equation}
where, for $m,n = 0, 1 . . .$, $l = 1, 2, . . .$,
\begin{eqnarray}
K_{mnl}(x,y,z)&=&\frac{H_n[\frac{\sqrt{2}y}{w(x)}]H_m[\frac{\sqrt{2}z}{w(x)}]{\mathrm{exp}}[-\frac{z^2+y^2}{w^2(x)}]}{w(x)\sqrt{\pi 2^{n+m-2}m!n!L}},\\
\psi_{mnl}(x,y,z)&=&kx-\phi(x)(m+n+1)+k\frac{z^2+y^2}{2R(x)},
\label{factor}
\end{eqnarray}
Here $g_0=\mu\sqrt{\omega_c/\epsilon_0 V}$, $\epsilon_0$ is the vacuum permittivity, $V$ shows the
volume of the cavity and $\mu$ is the electric-dipole transition matrix element. $H_n(y)$ is the $n$-th Hermite
polynomial, $w(x)=w_0\Big[1+(\frac{x}{x_R})^2\Big]^{\frac{1}{2}}$ is the beam waist at $x$ which is defined as the distance out from the axis
center of the beam where the irradiance drops to $1/e^2$ of its values on axis, $R(x)=x+x^2_R/x$  is the radius
of curvature of the wavefront at $x$, $\phi(x) = {\mathrm{arctan}}(x/x_R)$ is the Gouy phase shift\cite{Siegman}, $w_0$
is the cavity waist radius which depends on the geometry
of the Fabry-Perot cavity and $x_R=w_0^2k/2$ is the Rayleigh range which combines the wavelength and waist radius into a
single parameter and completely describes the divergence of the Gaussian beam. Note that the Rayleigh range is the
distance from the beam waist to the point at which the beam radius has increased to $\sqrt{2}w_0$. The coupling rate $\chi_{mnl}(\vec r_0,x)$ depends on the initial atomic position $\vec r_0$(measured from the cavity waist) as well as the displacement $x=x_{\mathrm{ZPF}}(b+b^{\dagger})$ of the mirror due
to $k = \omega_{\mathrm{eff}}(x)/c$, where $\omega_{\mathrm{eff}}(x)=\omega_{\mathrm{c}}(1-\frac{x}{L})$. As we will see in the next section, this dependence to the position of MR is responsible for the appearance of a new type of optomechanical nonlinearity.
Finally, the Hamiltonian $H_{\mathrm{dri}}$ describes the input driving by a laser with frequency $\omega_{l}$ and
amplitude $|\mathcal{E}|=\sqrt{2 \kappa P/\hbar \omega_l}$, where $P$ is the input laser power and $\kappa$
is the cavity loss rate through its input port.
\subsection{The nonlinear atom-field-mirror coupling}
As we have seen the Gaussian standing-wave structure
of the cavity mode leads to the field-atom coupling rate $\chi_{mnl}(r_0,x)$. Such field-atom coupling in the presence of the mode structure of the field has been studied extensively in the literature and it has been shown that a certain type of nonlinearity is prepared in the field-atom system. For instance, in Refs.\cite{joshi,joshi2} the influences of the atomic motion and the field-mode structure on the atomic dynamics have been investigated. It has been shown that the atomic motion and the field-mode structure give rise to nonlinear transient effects in the atomic population which are similar to self-induced transparency and adiabatic effects. In our treatment, the spatial field-mode structure leads to the appearance of an intensity-dependent interaction among the intracavity optical
mode, the MR and the single atom. To show this, we assume that the atom is well located at the transverse (polar) coordinate(measured
from the cylindrically symmetric cavity axis along the $x$ direction)
$\rho_0=\sqrt{x_0^2+y_0^2}=\mu\, w(x_0)$ where $0\leq \mu\leq1$. In the $x$ direction the localization of the atom can be expressed as $k_0x_0=\epsilon\pi$ for $\epsilon>0$,  where $k_0=\omega_c/c$.

At lowest order of the optical modes, i.e., $m=n=0,l=1$, the tripartite coupling rate reduces to
\begin{equation}\label{coupling3}
\chi_{001}(\vec r_0,x)\equiv\chi(x)=\frac{2g_0}{e^\mu w(x_0)\sqrt{\pi L}}{\mathrm{sin}}\Big[kx_0-\phi(x_0)-\frac{\pi}{2}+\frac{2\mu x_0}{k w^2_0}\Big],
\end{equation}
which can be rewritten in terms of the mirror position by using the position dependence of the wavelength $k=k_0(1-x/L)$ as
\begin{equation}\label{couplinga}
\chi(x)=\frac{2g_0}{e^{\mu} w(x_0)\sqrt{\pi L}}{\mathrm{sin}}\Big[\theta+\eta_0x\Big],
\end{equation}
where $\eta_0=\frac{2\mu x_0}{w^2_0k_0L}$ and
\begin{equation}\label{coupling234}
\theta=(1+\frac{2\mu}{w^2_0k^2_0})k_0x_0-\phi(x_0)-\frac{\pi}{2}.
\end{equation}
By substituting $x=x_{\mathrm{ZPF}}(b+b^{\dagger})$ in Eq.(\ref{couplinga}) we obtain
\begin{equation}\label{couplinga22}
\chi(b,b^{\dagger})=\frac{g_0}{i e^\mu w(x_0)\sqrt{\pi L}}\Big\{e^{i\theta}{\mathrm{exp}}[i\eta(b+b^{\dagger})]-h.c\Big\},
\end{equation}
where the parameter
\begin{eqnarray}
\eta=\eta_0x_{\mathrm{ZPF}}=\frac{2\pi\mu \epsilon}{w^2_0k^2_0L}\sqrt{\frac{\hbar}{m\omega_m}},
\end{eqnarray}
is the so-called "Lamb-Dicke parameter"(LDP). By using the Baker-Campbell-Hausdorff theorem in
Eq.(\ref{couplinga22}) and expanding the exponential terms in terms of $b$ and $b^{\dagger}$, the coupling rate
can be written as
\begin{equation}\label{couplinga3}
\chi(b,b^{\dagger})=\frac{g_0e^{-\eta^2/2}}{i e^\mu w(x_0)\sqrt{\pi L}}\Big\{e^{i\theta}\sum_{m,m'}\frac{(i\eta b^{\dagger})^{m'}(i\eta b)^{m}}{m!m'!}-h.c\Big\}.
\end{equation}
By using the bosonic commutation relation of the operators $b$ and $b^{\dagger}$, the \textit{j}-th term of the field-atom coupling rate is obtained as follows
\begin{eqnarray}\label{coupling5}
\chi_j(b,b^{\dagger})&&= \frac{g_0e^{-\eta^2/2}}{i e^{\mu} w(x_0)\sqrt{\pi L}}
\Big[e^{i\theta}\sum_{m}\frac{(i\eta)^{2m+j}(b^{\dagger})^{m+j} b^m}{m!(m+j)!}-h.c\Big]\nonumber\\
&&=g_{j,\mu}(b^{\dagger})^jf_j(n_b)+h.c,
\end{eqnarray}
where $g_{j,\mu}=\frac{g_0e^{-\eta^2/2}(i\eta)^{j}}{i e^{\mu} w(x_0)\sqrt{\pi L}}e^{i\theta}$
describes the effective atom-field-mirror coupling rate and the Hermitian nonlinearity function $f_j(n_b)$ defined by
\begin{equation}\label{non}
f_j(n_b)=\sum_{m}\frac{(i\eta)^{2m}n_b!}{m!(m+j)!(n_b-m)!}=\frac{n_b!}{(n_b+j)!}L_{n_b}^j(-\eta^2),
\end{equation}
with $n_b=b^{\dagger}b$ and $L_{n_b}^j(-\eta^2)$ as the associated Laguerre polynomial,
describes a nonlinear atom-field-mirror coupling through \textit{j}-phonon excitations of
the vibrational sideband. The nonlinearity function $f_j(n_b)$
has a central role in our treatment. It determines the form of nonlinearity of
the intensity dependence of the coupling among the cavity
field, the MR and the single atom. As we will see, this function drastically influences the
dynamics of the system, its entanglement properties and its responsible for the appearance of NMS with high
visibility in the displacement spectrum of the MR. Fig.2(a) shows the nonlinearity function $f_j(n_b)$ as a
function of $n_b$ and for different values of phonon excitation
number $j$. As is seen, this function has maximum contribution
around small values of vibrational acoustic excitation $n_b$. Furthermore, by increasing the number $j$ the strength of the nonlinearity function $f_j(n_b)$ decreases considerably. On the other hand, Fig.2(b) shows that the nonlinearity decreases by increasing the LDP. It is remarkable that, for the higher orders of the vibrational sideband, $j\geq3$,  this function is not any more sensitive to the LDP.
\begin{figure}[ht]
\centering
\includegraphics[width=3.in]{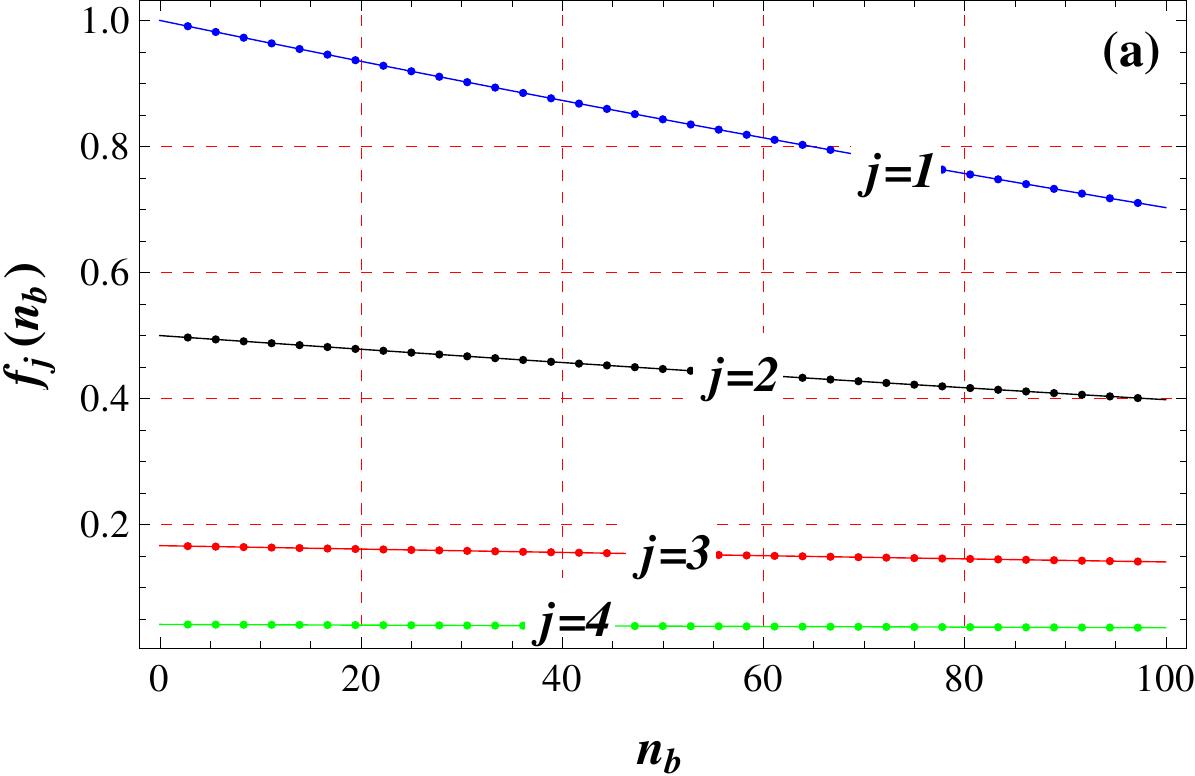}
\includegraphics[width=3.in]{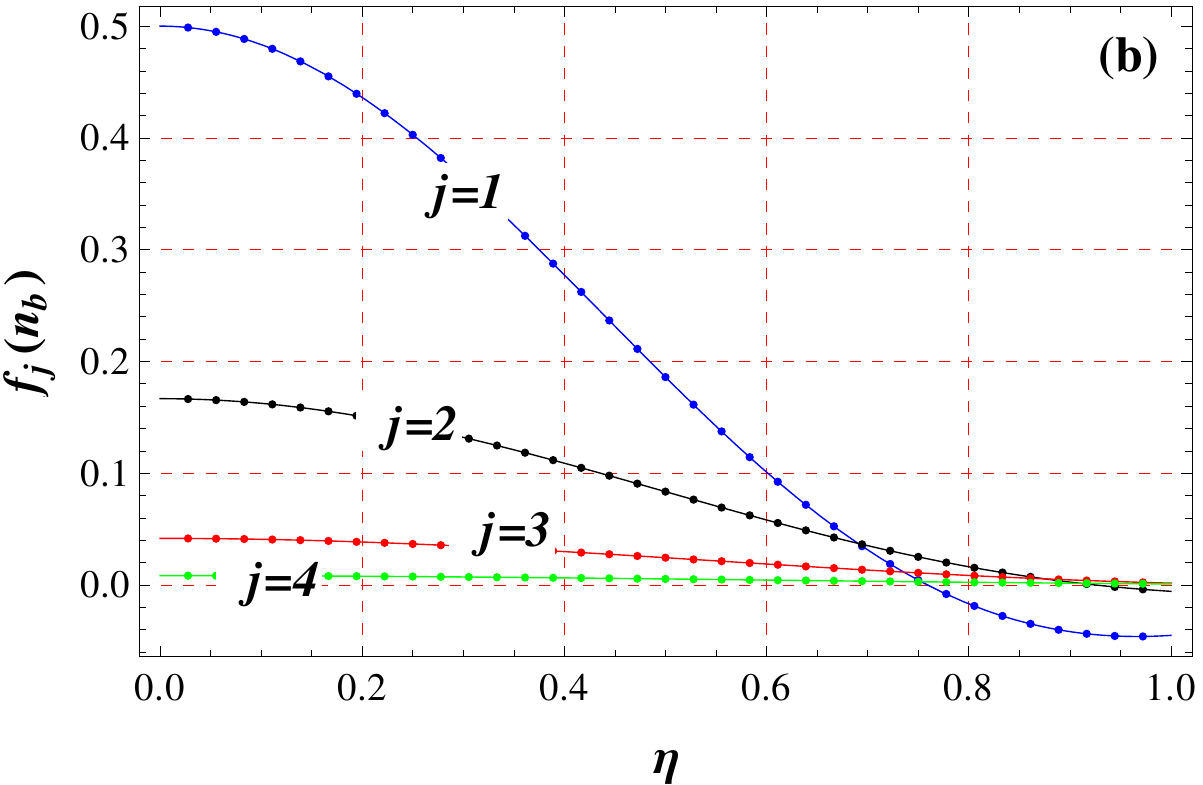}
\label{fig:entdetuning4}
\caption{
The nonlinearity function $f_j(n_b)$ as a function of:(a) phonon number $n_b$  for $\eta=0.08$ and for different values of vibrational sideband $j$ and (b) the Lamb-Dicke parameter, $\eta$, for $n_b=10$ and for different values for vibrational sideband $j$. }
\end{figure}

Now, by substituting Eq.(\ref{coupling5}) into the interaction Hamiltonian of Eq. (\ref{hamiltonian1}), we obtain the nonlinear form of the Hamiltonian as
\begin{equation}
H_{\mathrm{int}}^{(j)}=-\hbar \xi_0 (b+b^{\dagger})a^{\dagger}a+\hbar \Big[g_{j,\mu}(b^{\dagger})^j f_j(n_b)+h.c\Big]\Big[a\sigma^{+}+h.c\Big].
\label{hamiltonian3}
\end{equation}
Near the \textit{photon-phonon} resonance \cite{Yue} where the frequencies satisfy
$\omega_{\mathrm{m}}+\omega_{\mathrm{e}}-\omega_{\mathrm{c}}\simeq 0$, the rotating-wave approximation reduces the above Hamiltonian to
\begin{equation}
H_{\mathrm{int}}^{(j)}\simeq-\hbar \xi_0(b+b^{\dagger})a^{\dagger}a+\hbar \Big[g_{j,\mu}(b^{\dagger})^j f_j(n_b)a\sigma^{+}+h.c\Big].
\label{hamiltonian4}
\end{equation}
This Hamiltonian describes a nonlinear tripartite atom-field-mirror coupling and represents a novel type of optomechanical intensity-dependent interaction. The Hamiltonian (\ref{hamiltonian4})
is general and one can recover the results of Ref.\cite{genes} by taking $j=0$,  $f_j(n_b)\rightarrow1$ or results of Refs.\cite{Yue} and \cite{wang} by setting $j=1$, $f_j(n_b)\rightarrow1$. To study the system dynamics we restrict our
investigation by considering the first excitation of the vibrational sideband i.e., $j=1$. In this limit one may
use the following simple form of the Hamiltonian
\begin{equation}
H_{\mathrm{imt}}^{(j=1)}\simeq-\hbar \xi_0(b+b^\dagger)a^{\dagger}a+\hbar g_{\mu} \Big[b^{\dagger} f(n_b)a\sigma^{+}+\sigma^{-}a^{\dagger} f(n_b)b\Big],
\label{hamiltonian5}
\end{equation}
where $f(n_b)\equiv f_1(n_b)$ and $g_{\mu}=\frac{g_0 e^{-\eta^2/2}\eta}{e^{\mu} w(x_0)\sqrt{\pi L}}$. Since we deal with a well localized atom we can assume $\theta=\pi$ in Eq.(\ref{coupling234}), which is
realized by choosing a proper value of $\epsilon$ corresponding to the position of atom in the $x$ direction.
We pointed out that for the experimentally feasible parameters of
the system under consideration\cite{anet}, i.e., $k_0\simeq10^{6}\mathrm{m^{-1}}$, $m=10$pg, $\omega_m/2\pi=10$MHz, $L=1\mu $m and for the smallest achievable value of the cavity beam waist
$w_0\geq10^{-9}\mathrm{m}$ the LDP is always less than one, i.e., $\eta<1$. In this limit we can keep terms up to first order in
the phonon number $n_b$, and safely truncate the summation of Eq.(\ref{non}),
\begin{equation}
f_{j=1}(n_b)\simeq1-\frac{\eta^2}{2}n_b.
\label{non1}
\end{equation}
By substituting Eq.(\ref{non1}) into the Hamiltonian (\ref{hamiltonian5}) one can write the interaction of Hamiltonian as
\begin{eqnarray}
H_{\mathrm{int}}&=-\hbar \xi_0(b+b^{\dagger})a^{\dagger}a+\hbar g_{\mu} \Big[b^{\dagger}a\sigma^{+}+\sigma^{-}a^{\dagger}b\Big]-\nonumber\\
&-\frac{\hbar\eta^2 g_{\mu}}{2} \Big[b^{\dagger}n_b a\sigma^{+}+\sigma^{-}a^{\dagger}n_b b\Big].
\label{hamiltonian6}
\end{eqnarray}
Note that the first and second terms of the above Hamiltonian denote the standard tripartite atom-field-mirror
coupling which recently has been studied in Refs.\cite{Yue} and \cite{wang}. The third term denotes an intensity-dependent coupling
among the three subsystems of atom-field-mirror. This type of nonlinear coupling is attributed to the spatial field-mode structure at the position of the atom.

\section{dynamics of the system}

To describe the dynamical behavior of the system under consideration it is necessary to consider the fluctuation-dissipation processes affecting the three subsystems.
For this purpose, we first assume the excitation probability of the single atom to be small. In this limit, the dynamics
of the atomic polarization can be described in terms
of the bosonic operators $c$ and $c^{\dagger}$($[c,c^{\dagger}]=1$)~\cite{genes,Holstein}, where the atomic annihilation
operator is defined as $c=\sigma^-/\sqrt{|\langle\sigma^z\rangle|}$. This is valid in the
low atomic excitation limit, i.e., when the atom is initially
prepared in its ground state\cite{genes}. This means that the single-atom excitation probability should be much smaller than one. i.e., $ \frac{g_0|\alpha_s|^2}{\Delta_{a}^2+\gamma_a^2}\ll 1 $, where $\Delta_{a}=\omega_{\mathrm{a}}-\omega_l$ is the atomic detuning with respect to the laser and $\gamma_a$ is the decay rate of the excited atomic level. Therefore, the bosonization of the atomic operators is valid only if $ g_0\ll \Delta_{a}^2+\gamma_a^2$ that is the atom is weakly coupled to the cavity. 

The dynamics of the system is fully characterized by the following set of nonlinear quantum Langevin equations, written in the frame rotating at the input laser frequency,
\begin{subequations}
\begin{eqnarray}
\dot{c}&=&-[\gamma_a+\mathrm{i}\Delta_{a}]c -\mathrm{i} G(1-\frac{\eta^2}{2}n_b)ab^{\dagger}+\sqrt{2\gamma_a} F_{\mathrm{a}},\\
\dot{a}&=&-[\kappa+\mathrm{i}\Delta_{0f}]a +\rmi \xi_0 a(b+b^{\dagger})-\mathrm{i} G(1-\frac{\eta^2}{2}n_b)bc+\nonumber\\
&&\,\,\,\,\,\,\,\,\,\,\,\,\,\,\,\,\,\,\,\,\,\,\,\,\,\,\,\,+\mathcal{E}+\sqrt{2\kappa} a_{\mathrm{in}},\\
\dot{b}&=&-[\gamma_m+\rmi\omega_m]b +\rmi \xi_0 a^{\dagger}a-\rmi G[(1-\eta^2n_b)ac^{\dagger}-\frac{\eta^2}{2}a^{\dagger}cb^2]+\nonumber\\
&&\,\,\,\,\,\,\,\,\,\,\,\,\,\,\,\,\,\,\,\,\,\,\,\,\,\,\,\,\,+\sqrt{2\gamma_m} b_{\rmin},
\end{eqnarray}
\label{nonlinear}
\end{subequations}
where $\Delta_{0f}=\omega_{\mathrm{c}}-\omega_l$ is the cavity detuning with respect to the laser, $G=g_\mu\sqrt{|\langle\sigma^{\mathrm{z}}\rangle|}$ and $\gamma_m$ is the decay rate of the vibrational mode of the MR. The motional quantum fluctuation $b_{in}(t)$ satisfies the following relations\cite{gardiner}
\begin{eqnarray}\label{correfield1}
\langle b_{in}(t)b_{in}^{\dagger}(t')\rangle&=&[\langle n_{b,th}\rangle+1]\delta(t-t'),\nonumber\\
\langle b_{in}^{\dagger}(t)b_{in}(t')\rangle&=&\langle n_{b,th}\rangle\delta(t-t'),\\
\langle b_{in}(t)b_{in}(t')\rangle&=&\langle b_{in}^{\dagger}(t)b_{in}^{\dagger}(t')\rangle=0,\nonumber
\end{eqnarray}
where $\langle n_{b,th}\rangle$ is the mean number of phonons in the absence of optomechanical coupling, determined
by the temperature of the mechanical bath $T$,
\begin{eqnarray}
\langle n_{b,th}\rangle=\frac{1}{e^{\frac{\hbar \omega_m}{k_B T}}-1}.
\end{eqnarray}
The only nonvanishing
correlation function of the noises affecting the atom and the
cavity field is $\langle a_{in}(t)a_{in}^{\dagger}(t')\rangle=\langle F_{a}(t)F_{a}^{\dagger}(t')\rangle=\delta(t-t')$\cite{gardiner}.
\subsection{Linearization of QLEs}
Our aim is to study the conditions under which one can efficiently correlate and entangle the atom and the mechanical resonator
by means of the common interaction with the intracavity optical mode. As shown in Refs.\cite{Genes2008} and
\cite{glave}, a straightforward way for achieving stationary and robust entanglement
in continuous variable optomechanical systems, is to choose an operating point where the cavity is
intensely driven so that the intracavity field is strong, which is realized for high-finesse
cavities and enough driving power. Therefore, we focus onto
the dynamics of the fluctuations around the classical steady
state by decomposing each operator in Eqs.~(\ref{nonlinear}) as the sum of its steady-state value and a small fluctuation, e.g., $a=\alpha_s+\delta a$, $b=\beta_s+\delta b$, and $c=c_s+\delta c$. The steady state terms of these operators are given by
\begin{subequations}
\begin{eqnarray}
b_s&=& \frac{\alpha_s(\xi/2-G_3)}{(\omega_m-\rmi \gamma_m)},\\
c_s&=&\frac{G_2\alpha_s}{(\rmi\gamma_a-\Delta_{a})},\\
\mathcal{E}&=&\alpha_s\Big[\rmi\Delta_{f}+\kappa-\frac{|G_2|^2}{(\gamma_a+\rmi \Delta_{a})}\Big],
\end{eqnarray}
\label{nonlinear222}
\end{subequations}
where $\Delta_f=\Delta_{0f}-2\xi_0Re(b_s)$ denotes the effective optomechanical detuning and $ \xi =2\xi_0a_s $. The other parameters are defined in the appendix. In the linearization manner, we also obtain the following linear QLEs for the quantum fluctuations of the triple system
\begin{subequations}
\begin{eqnarray}
\delta\dot{c}&=&-[\gamma_a+\rmi\Delta_{a}]\delta c -\rmi G\Big[(1-\frac{\eta^2}{2}|b_s^2|)(a_s \delta b^{\dagger}+b^*_{s}\delta a)\nonumber\\
&-&\frac{\eta^2}{2}a_s b^*_s(b_s\delta b^{\dagger}+b^*_s\delta b)\Big]+\sqrt{2\gamma_a} F_{\rm a},\\
\delta\dot{a}&=&-[\kappa+\rmi\Delta_{0f}]\delta a +\rmi \xi_0\Big[\delta a(b_s+b^*_s)+a_s(\delta b+\delta b^{\dagger})\Big]\nonumber\\
&-&{\rm i} G\Big[(1-\frac{\eta^2}{2}|b|^2)(c_s\delta b+b_s \delta c)-\frac{\eta^2}{2}b_s c_s(b_s \delta b^{\dagger}+b^*_s\delta b)\Big]\nonumber\\
&+&\sqrt{2\kappa} a_{\rmin},\\
\delta\dot{b}&=&-[\gamma_m+\rmi\omega_m]\delta b +\rmi \xi_0 a_s(\delta a^{\dagger}+\delta a)\nonumber\\
&-&\rmi G\Big[(1-\eta^2|b_s|^2)(a_s \delta c^{\dagger}+c^*_s\delta a)-\eta^2 a_s c^*_s(b_s \delta b^{\dagger}+b^*_s\delta b)\nonumber\\
&-&\frac{\eta^2}{2}\{b^2_s(c_s \delta a^{\dagger}+a_s \delta c)+2a_s b_s c_s \delta b\}\Big]+\sqrt{2\gamma_m} b_{\rmin},
\end{eqnarray}
\label{nonlinear2}
\end{subequations}
in terms of the fluctuations of the quadrature operators,
\begin{eqnarray}\label{eqmotion1}
\delta X_a&&=\frac{1}{\sqrt{2}}(\delta a+\delta a^{\dagger}),
\delta Y_a=\frac{1}{\sqrt{2}\rmi}(\delta a-\delta a^{\dagger}),\\
\delta X_c&&=\frac{1}{\sqrt{2}}(\delta c+\delta c^{\dagger}),
\delta Y_c=\frac{1}{\sqrt{2}\rmi}(\delta c-\delta c^{\dagger}),\\
\delta q&&=\frac{1}{\sqrt{2}}(\delta b+\delta b^{\dagger}),
\delta p=\frac{1}{\sqrt{2}\rmi}(\delta b-\delta b^{\dagger}).
\end{eqnarray}
The resulting linearized QLEs can be written in the following compact matrix form
\begin{equation}
\dot{u}(t)=A u(t)+n(t),
\label{compact}
\end{equation}
where $u(t) =\big[\delta q(t), \delta p(t),\delta X_a(t), \delta Y_a(t),\delta X_c(t),\delta Y_c(t)\big]^{\mathsf{T}}$
 is the vector of CV fluctuation operators and
 $n(t) =[\sqrt{2\gamma_m}q^{\rmin}(t), \sqrt{2\gamma_m}p^{\rmin}(t),\sqrt{2\kappa}X_a^{\rmin}(t), \sqrt{2\kappa}Y_a^{\rmin}(t),\\
  \sqrt{2\gamma_a}X_c^{\rmin}(t), \sqrt{2\gamma_a}Y_c^{\rmin}(t)]^{\mathsf{T}}$
is the corresponding vector of noises.
Moreover, the drift matrix $A$ is a $6 \times 6$ matrix
\begin{equation}
A=\left(\begin{array}{cccccc}
    -\Gamma_{1\rmm} & \Omega_{1\rmm} & -M^I_2 & M^R_2& -M^I_1 & M^R_1 \\
    -\Omega_{2\rmm} & -\Gamma_{2\rmm} &-M^R_2 & -M^I_2 & -M^R_3 & -M^I_3 \\
    -G^I_1 & G^R_1 & -\kappa & \Delta_f & -G^I_2 & G^R_2 \\
    \xi- G^R_3 & - G^I_3 & -\Delta_f & -\kappa &  -G^R_2 & -G^I_2\\

    -N^I_2& N^R_2 &  -N^I_1 & N^R_1 & -\gamma_a & \Delta_{a} \\
    -N^R_3 & -N^I_3 & -N^R_1 & -N^I_1 & -\Delta_{a} &-\gamma_a
  \end{array}\right),
\label{drift}
\end{equation}
where $O_i^R$ and $O_i^I$ denote the real and imaginary parts of parameter $O_i$, respectively. The other matrix elements are defined in the appendix.
%
\subsection{Stationary quantum fluctuations}
Here, we focus our attention on the stationary properties of the system. For this purpose we should consider the
steady state condition governed by Eq. (\ref{compact}). The steady state is reached when
the system is stable, which occurs if and only if all the
eigenvalues of the matrix $A$ have a negative real part. These stability
conditions can be obtained, for example, by using the Routh-Hurwitz criterion\cite{rh}.

The steady state is a zero-mean Gaussian state due to the
fact that the dynamics of the fluctuations is linearized and all
noises are Gaussian. As a consequence, it is fully characterized
by the $6\times6$ stationary correlation matrix (CM) V, with matrix
elements

\begin{equation}
 V_{ij}=\frac{\langle u_i(\infty)u_j(\infty)+ u_j(\infty)u_i(\infty)\rangle}{2}.
\label{CM}
\end{equation}
The formal solution of Eq.(\ref{compact}) yields\cite{Genes2008}
\begin{equation}
V_{ij}=\int_0^{\infty} ds \int_0^{\infty}ds' M_{ik}(s) M_{jl}(s')D_{kl}(s-s'),
\label{vij}
\end{equation}
where $M(t)=\exp(A t)$ and $D(s-s')$ is the diffusion matrix,
the matrix of noise correlations, defined as $D_{kl}(s-s')=\langle n_k(s)n_l(s')+n_l(s')n_k(s) \rangle/2$.
For the noise diffusion matrix we have $D(s-s')=D\delta(s-s')$, where
$D=\mathrm{diag}[\gamma_m(2\bar{n}_b+1), \gamma_m(2\bar{n}_b+1),\kappa,\kappa,\gamma_a,\gamma_a]$. Therefore, Eq.(\ref{vij}) is simplified to
\begin{equation}
V =\int_0^{\infty} d \omega V(\omega),
\label{cm3}
\end{equation}
where
\begin{equation}
V(\omega)= M(\omega)D M(\omega)^{\mathsf{T}}.
\label{vs}
\end{equation}
When the stability conditions are satisfied ($M(\infty)=0$), one can obtain the following Lyapunov equation
\begin{equation}
AV+VA^{\mathsf{T}}=-D.
\label{lyap}
\end{equation}
Equation (\ref{lyap}) is a linear equation for $V$ and can be straightforwardly
solved. However, the explicit form of $ V $ is complicate and
will not be reported here.
\section{entanglement properties of the steady-state of the tripartite system}
In this section we examine the entanglement properties of
the steady state of the tripartite system under consideration. For this purpose, we consider the
entanglement of the three possible bipartite subsystems that can be obtained by tracing over the remaining
degrees of freedom. Such bipartite entanglement will be quantified by using the logarithmic negativity \cite{eis},
\begin{equation}\label{en}
E_N=\mathrm{max}[0,-\mathrm{ln} 2 \eta^-],
\end{equation}
where $\eta^{-}\equiv2^{-1/2}\left[\Sigma(V_{bp})-\sqrt{\Sigma(V_{bp})^2-4 \mathrm{det} V_{bp}}\right]^{1/2}$  is the lowest symplectic eigenvalue of the partial transpose of the $4 \times 4$ CM, $V_{bp}$, associated with the selected bipartition, obtained by neglecting the rows and columns of the uninteresting mode,

\begin{equation}\label{loga}
V_{bp}=\left(
     \begin{array}{cc}
       B & C \\
       C^T & B' \\
     \end{array}
   \right),
\end{equation}
and $\Sigma(V_{bp})\equiv \mathrm{det} B+\mathrm{det} B'-2\mathrm{det} C$.
\begin{figure}[ht]
\centering
\includegraphics[width=3in]{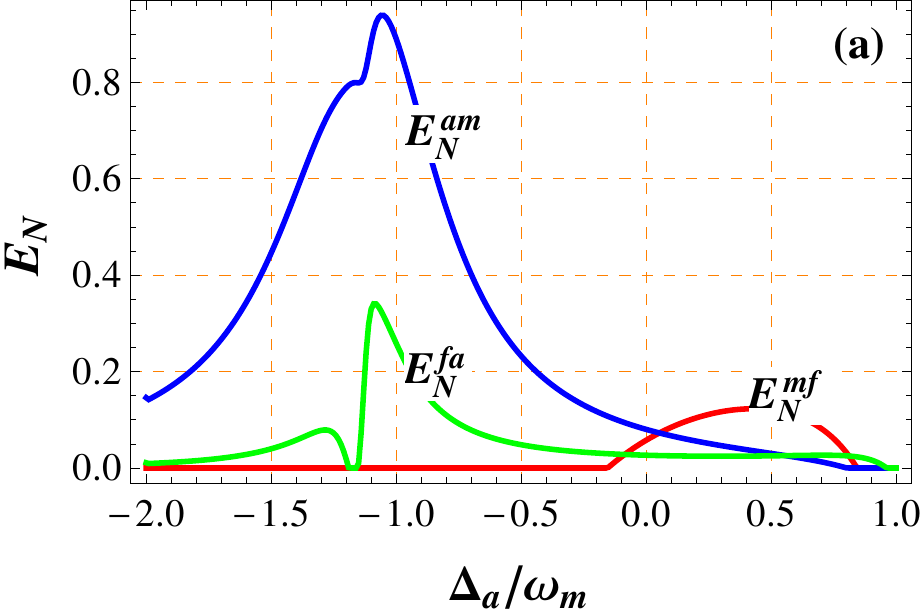}
\includegraphics[width=3in]{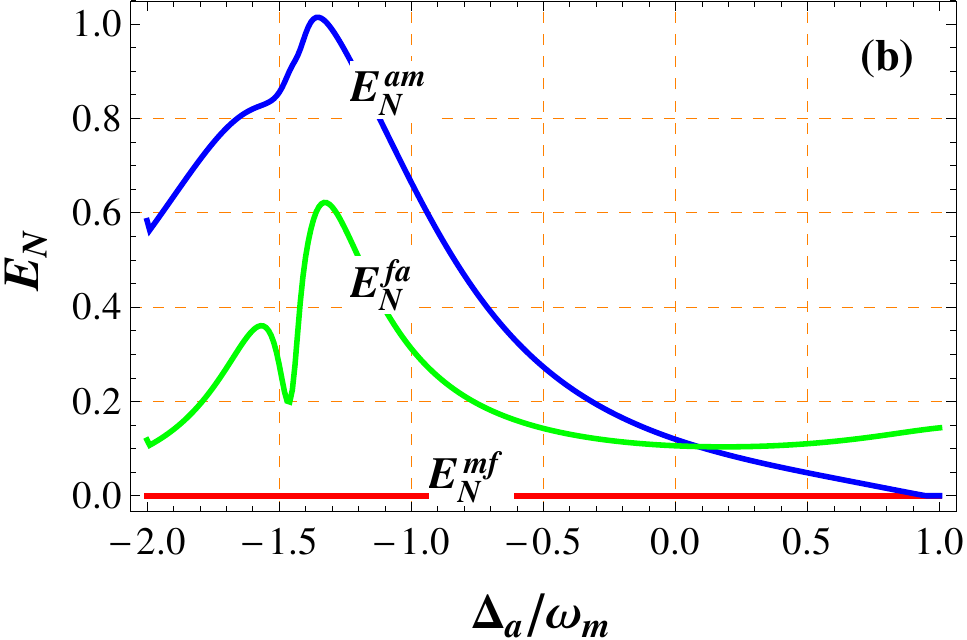}
\label{fig:entdetuning1}
\caption{
(Color online) Plot of $E_{N}$ of the three bipartite subsystems [$E_N^{am}$ (atom-mirror), $E_N^{fa}$ (field-atom), $E_N^{mf}$ (mirror-field)]versus the normalized atomic detuning $\Delta_a/\omega_{m}$ at fixed temperature $T=0.4$ K, and for two different values of the LDP: (a) $\eta=0.04$ and (b)$\eta=0.08$ . The optical cavity detuning has been fixed at $\Delta_f = -\omega_m$, while the other parameters are $\omega_m/2\pi=10$ MHz, $Q=11\times 10^5$, $m = 10$ pg, $\kappa=0.07\omega_m$, $k_{0}=10^6 m^{-1}$, $P_c=800\mu W$, $\gamma_a/2\pi=0.04\omega_m$, and $g_0/2\pi=10^3$Hz.}
\end{figure}

In Fig.3 we have plotted the three bipartite logarithmic negativities, $E_N^{am}$ (atom-mirror), $E_N^{fa}$ (field-atom), and $E_N^{mf}$ (mirror-field) versus the normalized atomic detuning $\Delta_a/\omega_{m}$ at fixed temperature $T=0.4$ K, for two different values of the LDP [ $\eta=0.04$, Fig.3a and $\eta=0.08$, Fig.3b] and for the experimentally feasible parameters\cite{anet}, i.e., a mechanical resonator with oscillation frequency $\omega_m/2\pi=10$ MHz,  quality factor $Q=11\times 10^5$, $m = 10$ pg and an optical cavity with length $L=1$ $\mu$m and damping rate $\kappa=0.07\omega_m$ driven by a laser with $k_{0}\simeq10^6 m^{-1}$  and power $P_c=800\mu W.$ The atom damping constant has been taken $\gamma_a/2\pi=0.04\omega_m$ with coupling constant $g_0/2\pi=10^3$Hz.

 The optical cavity detuning has been fixed at $\Delta_f = -\omega_m$ which turns out be the most convenient choice. As seen, by increasing LDP, the two bipartite entanglement of $E_N^{am}$ and $E_N^{fa}$ increase and the bipartite entanglement of $E_N^{mf}$ decreases overall. The reason is that, by increasing the LPD the tripartite atom-field-mirror coupling rate increases compared to the coupling rate of the bipartite field-mirror subsystem, or equivalently the parameter $G/\xi$ is increased. This result reveals that by changing the LDP one can control the tripartite coupling amplitude or even goes through the regime in which the tripartite system reduces to an effective bipartite subsystem.  However, the three logarithmic negativities do not behave in the same way and the entanglement sharing is evident. In particular, the entanglement of interest, i.e., $E_N^{am}$, increases at the expense of the mirror-field entanglement, while $E_N^{fa}$ remains always non-negligible.

\begin{figure}[ht]
\centering
\includegraphics[width=3in]{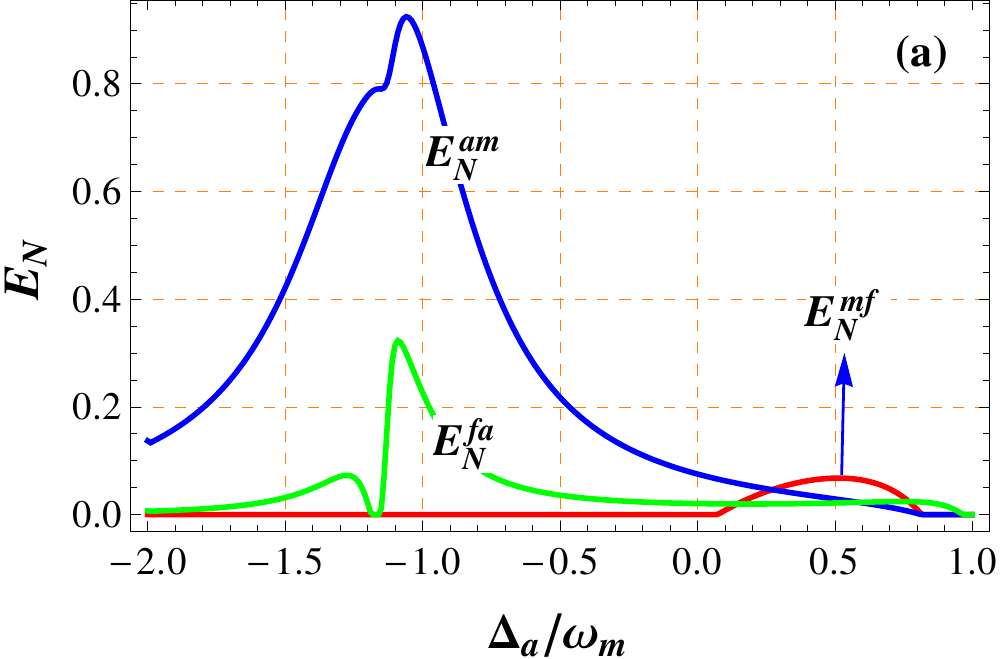}
\includegraphics[width=3in]{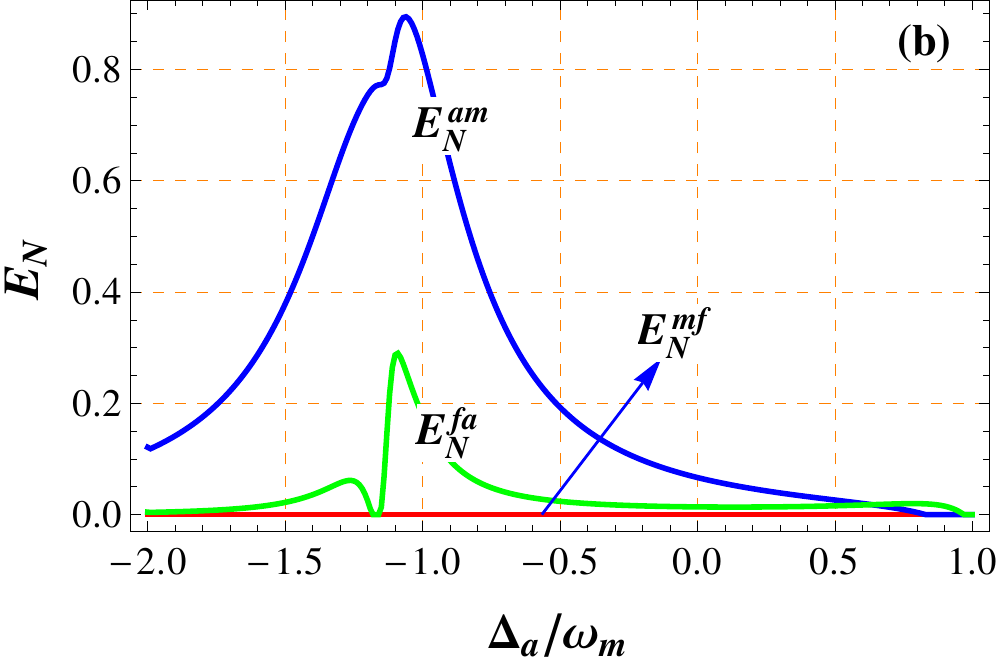}
\label{fig:entdetuning2}
\caption{
(Color online) Plot of $E_{N}$ of the three bipartite subsystems [$E_N^{am}$ (atom-mirror), $E_N^{fa}$ (field-atom), $E_N^{mf}$ (mirror-field)]versus the normalized atomic detuning $\Delta_a/\omega_{m}$ for a fixed value of the LDP, $\eta=0.04$, and for two different temperatures: (a)$T=1.2$K and (b) $T=3$K. The optical cavity detuning has been fixed at $\Delta_f = -\omega_m$ and
the other parameters are as in Fig. 3.}
\end{figure}

Fig.~4 shows the logarithmic-negativity of the three bipartite subsystems versus the normalized atomic detuning $\Delta_a/\omega_{m}$ for a fixed value of the LDP, $\eta=0.04$, and for two different temperatures: $T=1.2$ K (Fig.4a) and $T=3$ K (Fig.4b). The optical cavity detuning has been again fixed at $\Delta_f =- \omega_m$. As expected, the three kinds of bipartite entanglement decrease by increasing temperature, but the atom-mirror and field-atom entanglement show highly temperature robustness. However, the field-mirror entanglement shows extremely fragile entanglement robustness versus temperature and its logarithmic-negativity falls down to zero at $T=3$K.

Generally, the scheme is able to generate appreciable entanglement between the atom and the MR, especially by sharing from the mirror-field entanglement. Similar bipartite entanglement behavior can be observed in other similar tripartite systems, such as the atom-field-mirror scheme proposed in Ref.~\cite{genes},
the microwave-optical-mirror system of Ref.~\cite{shabir} and the two cavity optomechanical
setup of Ref.~\cite{pat}.
\begin{figure}[ht]
\centering
\includegraphics[width=3in]{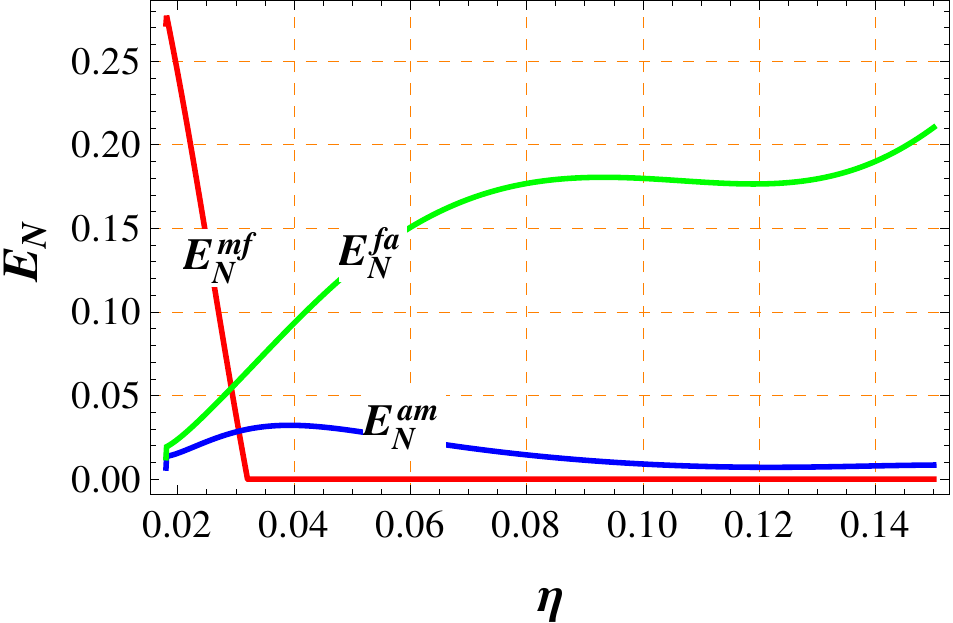}
\caption{
(Color online) Plot of $E_{N}$ of the three bipartite subsystems [$E_N^{am}$ (atom-mirror), $E_N^{fa}$ (field-atom), $E_N^{mf}$ (mirror-field)]versus the LDP, $\eta$, for fixed temperature $T=0.4$K. The optical cavity and atomic detuning are $\Delta_f = -\omega_m$ and $\Delta_a= \omega_m$, respectively. The other parameters are as in Fig. 3.}
\end{figure}

The effect of the LDP is also illustrated in Fig.~5, where we have plotted the logarithmic-negativity as a function of $\eta$ at fixed optical cavity detuning  $\Delta_f =- \omega_m$ and at atomic
detuning $\Delta_a=\omega_m$. It is clear that by increasing the LDP, the field-atom entanglement increases when the entanglement between the intracavity mode and the MR is drastically suppressed. We also observe that the atom-mirror logarithmic-negativity $E_N^{am}$ plateaus in a case as $\eta$ increases.
\begin{figure}[ht]
\centering
\includegraphics[width=3.4in]{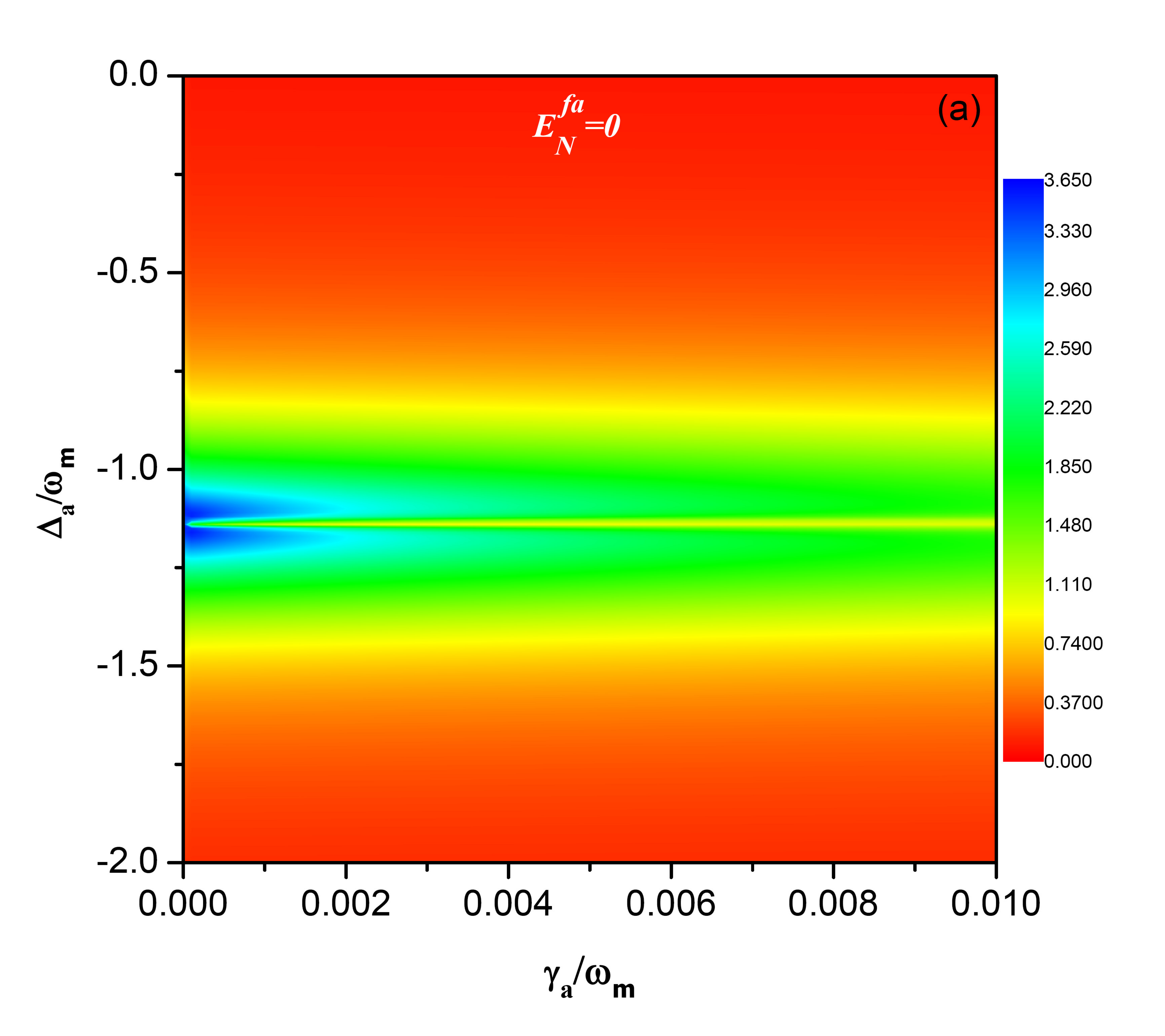}
\includegraphics[width=3.4in]{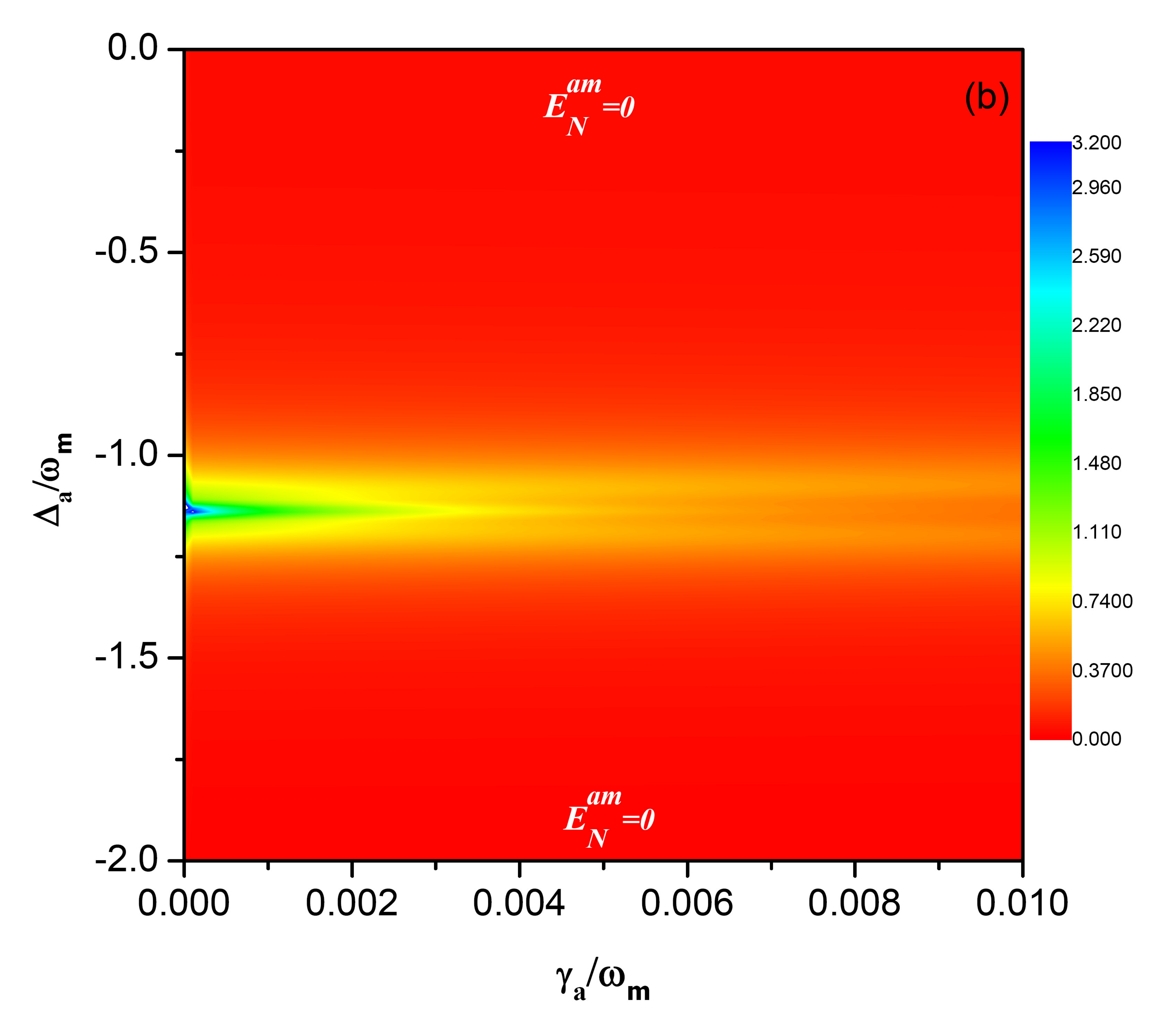}
\label{fig:entdetuning5}
\caption{
(Color online) Density plot of $E_{N}$ of the bipartite subsystems: (a)$E_N^{fa}$ and (b) $E_N^{am}$ versus $\Delta_a/\omega_{m}$ and $\gamma_a/\omega_{m}$ for $\eta=0.04$ and for $ T=0.4 $K. The optical cavity detuning has been fixed at $\Delta_f = -\omega_m$. The other parameters are as in Fig. 3.}
\end{figure}

Fig.6 shows $E_N^{fa}$ and $E_N^{am}$ versus $\Delta_a/\omega_{m}$ and $\gamma_a/\omega_{m}$ for $\eta=0.04$. As is clear, the $E_N^{fa}$ and $E_N^{am}$ are maximized around sideband $ \Delta_a\simeq\omega_{m}$. By increasing the atomic spontaneous emission rate $\gamma_a$ as we expected, both logarithmic-negativities decrease drastically.

The entanglement features of the tripartite system at the steady state
 can be observed by experimentally measuring the corresponding CM. This can be done by combining existing experimental techniques. By homodyning  the cavity output one can measure the cavity field quadratures.  Ref.\cite{vitali1} has proposed a scheme to measure mechanical position and momentum of the MR, in which by adjusting the detuning and bandwidth of an additional adjacent cavity, both position and momentum of the mirror can be measured by homodyning the output of the second cavity. Moreover,  by adopting the same scheme of Ref.\cite{pol}, the atomic polarization quadratures $X_a$ and $Y_a$ can be measured, i.e., by making a Stokes parameter measurement of a laser beam, shined transversal to the cavity and to the cell and off-resonantly tuned to another atomic transition. Very recently, Ref.\cite{pat2} has demonstrated the proof of principle of the use of a
Bose-Einstein condensate(BEC) as a diagnostic tool to determine the elusive mirror-light entanglement in a hybrid optomechanical device. In such a case, one dose find a working point such that the mirror-light entanglement is reproduced by the BEC- light quantum correlations.

\section{NORMAL-MODE SPLITTING in the displacement spectrum of the MR}

In this section, we show that the atom-field-mirror coupling leads to the
splitting of the normal mode into three modes [Normal Mode Splitting(NMS)]. The optomechanical NMS however involves driving four parametrically coupled nondegenerate modes out of equilibrium. The NMS does not appear in the steady state spectra but rather manifests itself in the fluctuation spectra of the mirror displacement. To study the NMS in our system we need to find out the displacement spectrum of mirror as:

\begin{eqnarray}\label{sq}
S_q(\omega)=\frac{1}{2\pi}\int d\Omega e^{-i(\omega+\Omega)t}&&\langle\delta q(\omega)\delta q(\Omega)+\delta q(\Omega)\delta q(\omega)\rangle \nonumber\\
&&=V_{11}(\omega),
\end{eqnarray}
where $V_{11}(\omega,\Omega)=1/2\langle\delta q(\omega)\delta q(\Omega)+\delta q(\Omega)\delta q(\omega)\rangle$ is an element of CM which is given by Eq.(\ref{vs}). Unfortunately, the analytical form of the displacement spectrum of the mirror is too complicate to put a clear physical interpretation on it. Thus, in the following, we give and analyze the results obtained by numerical calculations.
\begin{figure}[ht]
\centering
\includegraphics[width=3in]{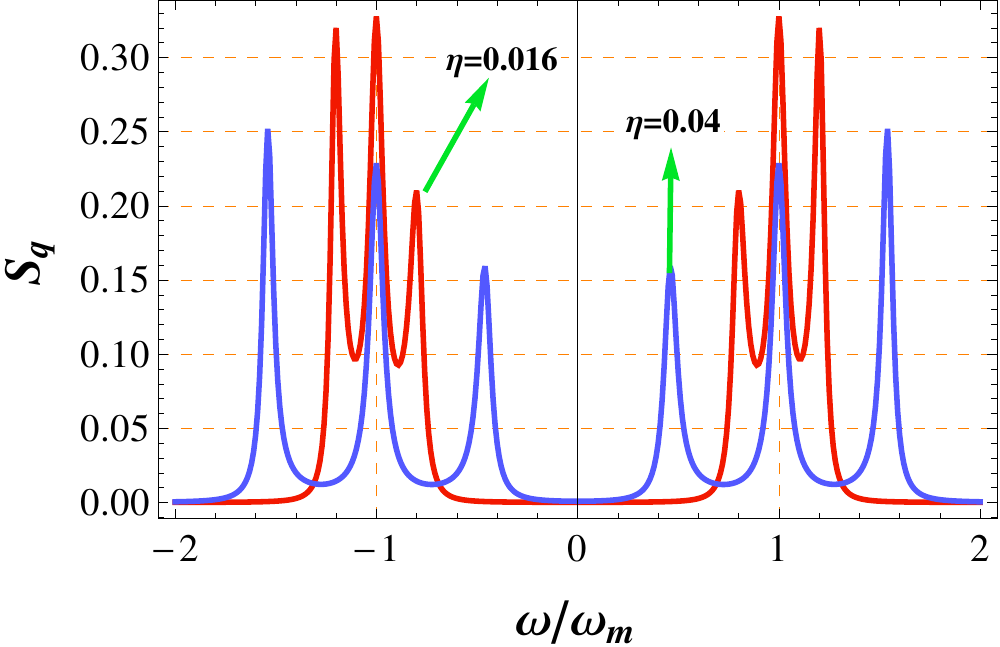}
\caption{(Color online) Normalized plot of the displacement spectrum $S_q(\omega)$ versus $\omega/\omega_m$ at fixed temperatures $T=0.4$K and for two different values of the LDP: $\eta=0.016$(red-line),  $\eta=0.04$(blue-line). The optical cavity and atomic detuning have been fixed at $\Delta_f =\omega_m$ and $\Delta_a= \omega_m$, respectively. The other parameters are as in Fig. 3.}
\end{figure}

Fig.~7 shows the displacement spectrum of the MR as a function of the normalized frequency $\omega/\omega_m$ at $\Delta_f=\omega_m$, $\Delta_a=\omega_m$ and for two different values of the LDP: $\eta=0.016$, $\eta=0.04$.  For the small values of the LDP, we observe the usual normal-mode splitting into two modes with central peaks at the sidebands $\omega=\pm \omega_m$. This figure shows a highly symmetric structure with respect to $\omega=0$. As is seen, by increasing the LDP the normal mode splits up into three
modes.

A more clear illustration of the three-mode splitting is shown in Fig.8. This figure shows the displacement spectrum of the MR versus the normalized frequency $\omega/\omega_m$ and atomic detuning $\Delta_a/\omega_m$ at $\Delta_f=\omega_m$. The three-mode splitting manifests itself mainly at $\Delta_a\simeq\omega$. By going through the region far from $\Delta_a\simeq\omega$, three-mode splitting merges into two-mode splitting around $\Delta_a\simeq0.75\omega_m$ and $\Delta_a\simeq1.35\omega_m$. The NMS is associated with the mixing among the vibrational mode of the MR, the fluctuations of the cavity field around the steady state and the fluctuations of the atomic mode. The origin of the
fluctuations of the cavity field is the beat of the pump photons with the photons scattered from the atom. For not so large values of the LDP(small nonlinearity) the field-atom coupling is much smaller than the field-mirror coupling. Therefore, the system simply reduces to the case of two mode coupling, i.e., coupling between
the mechanical mode and the photon fluctuations\cite{bhata}. When the LDP is large enough the mechanical
mode, the photon mode, and the atomic mode forms a
system of three coupled oscillators. The occurrence of splitting of the normal mode into three modes has been analyzed recently in another tripartite system, i.e., a cavity quantum optomechanical system of ultracold atoms in an optical lattice\cite{bhata}. Furthermore, similar three coupled oscillator experimental
results where two coupled cavities, each containing
three identical quantum wells\cite{sta} and one microcavity containing two quantum wells\cite{lin} have been reported.
\begin{figure}
\centering
\includegraphics[width=3.4in]{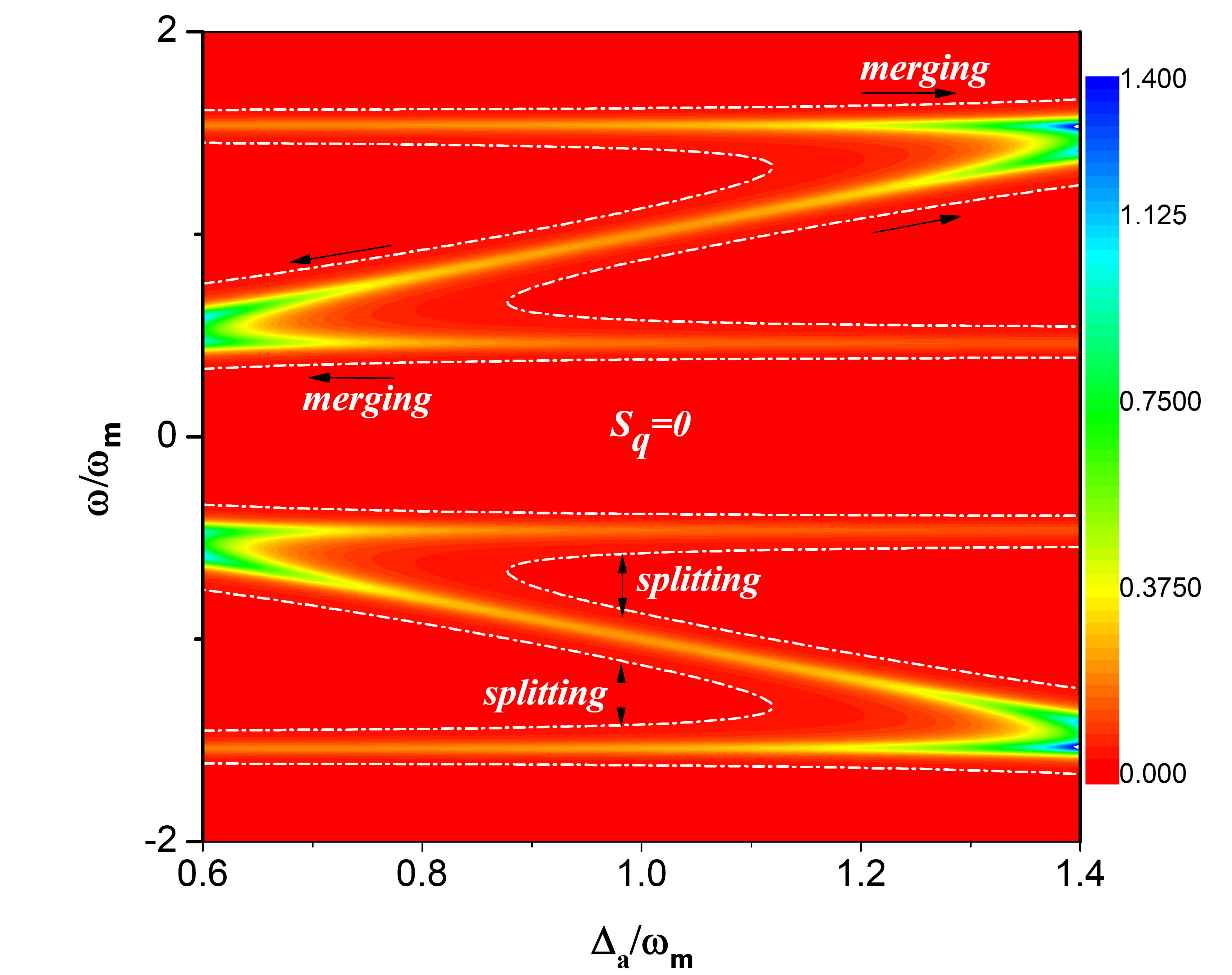}
\caption{(Color online) Density plot of the displacement spectrum $S_q(\omega)$ versus $\omega/\omega_m$ and  $\Delta_a/\omega_m$ for $T=0.4$K and $\eta=0.04$. The optical cavity detuning has been fixed at $\Delta_f =\omega_m$. The other parameters are as in Fig. 3.}
\end{figure}

It should be pointed out that to observe the NMS, the energy exchange between the three modes should take place on a time scale faster than the decoherence of each mode. The normal mode splitting into three modes due to local increasing of the LDP has also been reported in Ref.\cite{kli}, where the authors have shown that the NMS can be observed only if the coupling between the atoms and the cavity is strong enough. This strong coupling can be achieved by increasing the atom numbers. One experimental limitation could be spontaneous emission which leads to momentum diffusion and hence heating of the atomic sample\cite{mur}. In our model, we don't encounter such
a limitation and three-mode splitting is approached by proper choosing of the LDP.

\section{Conclusion}
In this work, we have proposed a theoretical scheme for the realization of tripartite intensity-dependent coupling among a single mode of a Fabry-Perot cavity with an oscillating mirror, a single two-level atom inside it, and a vibrational mode of the oscillating mirror. We have shown that in the presence of Gaussian standing-wave of the optical cavity mode, a type of tripartite between atom-mirror-field coupling can be manifested. To describe such interaction we then have found the general form of the corresponding nonlinear Hamiltonian. We have restricted our investigation to first vibrational sideband $j=1$ and have studied its dynamics by
adopting a QLE treatment. We have focused our attention on
the steady state of the system and in particular, on the
stationary quantum fluctuations of the system by solving
the linearized dynamics around the classical steady
state. We have seen that, in an experimentally accessible
parameter regime, the steady state of the system shows
both the tripartite and the bipartite CV entanglement.  We have shown that the LDP(as a measure of the strength of nonlinearity)can extremely modifies both the tripartite and the bipartite CV entanglement in the system. In particular, by increasing the LDP, one can see that the field-atom and atom-mirror entanglement increase at the expense of optical-mechanical entanglement.
The intracavity mode is able to mediate for the realization of a robust
stationary (i.e., persistent) entanglement between the MR mode
and the single two-level atom, which could be extremely
useful in quantum information/quantum computer networks in which
the MR modes are used for quantum communications\cite{manc1,manc2}, and
the atom is used as a qubit(e.g., solid-state qubits). Furthermore, we have analyzed the occurrence of the NMS in the displacement spectrum of the oscillating mirror. As we have shown, for a small value of the LDP, the usual normal-mode splitting into two modes with central peaks at the sidebands $\omega=\pm \omega_m$ is observed and by increasing the
LDP the normal mode splits up into three
modes. We have shown that, when the LDP is large enough the mechanical
mode, the photon mode, and the atomic mode forms a
system of three coupled oscillators. The realization of such a scheme
will also open new opportunities for the implementation
of quantum teleportation and/or the photon blockade process to prevent two or more photons from entering the cavity at the same time.

\section*{Acknowledgements}
The authors wish to thank The Office of Graduate
Studies of The University of Isfahan for their support.
\\
\\
\appendix*
\section{Definition of the elements of the drift matrix of Eq.(\ref{drift})}
In the drift matrix of Eq.(\ref{drift}) we have defined 
\begin{subequations}
\begin{eqnarray}
\Gamma_{1m}&=&\gamma_m+M_3^I, \Gamma_{2m}=\gamma_m+M_4^I,\nonumber\\
\Omega_{1m}&=&\omega_m+M_3^R, \Omega_{2m}=\omega_m+M_4^R,\nonumber\\
G_1&=&G\Big[c^*_s(1-\eta^2|b_s|^2)+\frac{\eta^2}{2}c_s b^2_s\Big],\nonumber\\
G_2&=&G b^*_s(1-\frac{\eta^2}{2}|b_s|^2),\nonumber\\
G_3&=&G\Big[c^*_s(1-\eta^2|b_s|^2)-\frac{\eta^2}{2}c_s b^2_s\Big],\nonumber\\
M_1&=&-G a_s\Big[ (1-\eta^2|b_s|^2)+\frac{\eta^2}{2}b_s^{2*}\Big],\nonumber\\
M_2&=&G a_s\Big[ (1-\eta^2|b_s|^2)-\frac{\eta^2}{2}b_s^{2*}\Big],\nonumber\\
M_3&=&G c_s\Big[(1-\eta^2|b_s|^2)+\frac{\eta^2}{2}b_s^2\Big],\nonumber\\
M_3&=&G c_s\Big[(1-\eta^2|b_s|^2)-\frac{\eta^2}{2}b_s^2\Big],\nonumber\\
M_4&=&-G\eta^2 a_s\Big[b_s^*c_s^*+c_s b_s-c_s^*b_s\Big],\nonumber\\
M_5&=&-G\eta^2 a_s\Big[b_s^*c_s^*+c_s b_s+c_s^*b_s\Big],\nonumber\\
N_1&=&b_s(1-\frac{\eta^2}{2}|b_s|^2),\nonumber\\
N_2&=&-a_s\Big[(1-\eta^2|b_s|^2)+\frac{\eta^2}{2}b_s^2\Big],\nonumber\\
N_3&=&a_s\Big[(1-\eta^2|b_s|^2)-\frac{\eta^2}{2}b_s^2\Big].\nonumber
\end{eqnarray}
\label{nonlinear234}
\end{subequations}

\bibliographystyle{apsrev4-1}

\end{document}